\documentclass[aps,prd,twocolumn,showpacs,amsmath,amssymb]{revtex4-2}
\usepackage{amsmath}
\usepackage{graphicx}
\usepackage{subfigure}
\usepackage{epstopdf}
\usepackage{color}
\usepackage{multirow}
\usepackage{setspace}
\usepackage{overpic}
\usepackage{amssymb}
\usepackage{braket}
\usepackage{makecell}
\usepackage[bookmarksnumbered, pdfstartview=FitH,colorlinks,urlcolor=blue, citecolor=blue,linkcolor=blue] {hyperref}
\usepackage{lineno}
\usepackage{bm}
\usepackage{rotating}
\usepackage[utf8]{inputenc}

\let\oldequation\equation
\let\oldendequation\endequation
\renewenvironment{equation}
 {\linenomathNonumbers\oldequation}
 {\oldendequation\endlinenomath}

\begin{document}
\title{\boldmath Partial wave analyses of $\psi(3686)\to p\bar{p}\pi^0$ and $\psi(3686)\to p\bar{p}\eta$}

\author{
	M.~Ablikim$^{1}$, M.~N.~Achasov$^{4,c}$, P.~Adlarson$^{76}$, O.~Afedulidis$^{3}$, X.~C.~Ai$^{81}$, R.~Aliberti$^{35}$, A.~Amoroso$^{75A,75C}$, Q.~An$^{72,58,a}$, Y.~Bai$^{57}$, O.~Bakina$^{36}$
% ...
\\
\vspace{0.2cm}
(BESIII Collaboration)\\
\vspace{0.2cm} {\it
$^{1}$ Institute of High Energy Physics, Beijing 100049, People's Republic of China\\
% ...
$^{3}$ Bochum  Ruhr-University, D-44780 Bochum, Germany\\
$^{4}$ Budker Institute of Nuclear Physics SB RAS (BINP), Novosibirsk 630090, Russia\\
% ...
$^{35}$ Johannes Gutenberg University of Mainz, Johann-Joachim-Becher-Weg 45, D-55099 Mainz, Germany\\
$^{36}$ Joint Institute for Nuclear Research, 141980 Dubna, Moscow region, Russia\\
% ...
$^{57}$ Southeast University, Nanjing 211100, People's Republic of China\\
$^{58}$ State Key Laboratory of Particle Detection and Electronics, Beijing 100049, Hefei 230026, People's Republic of China\\
% ...
$^{72}$ University of Science and Technology of China, Hefei 230026, People's Republic of China\\
% ...
$^{75}$ University of Turin and INFN, (A)University of Turin, I-10125, Turin, Italy; (B)University of Eastern Piedmont, I-15121, Alessandria, Italy; (C)INFN, I-10125, Turin, Italy\\
$^{76}$ Uppsala University, Box 516, SE-75120 Uppsala, Sweden\\
% ...
$^{81}$ Zhengzhou University, Zhengzhou 450001, People's Republic of China\\
\vspace{0.2cm}
$^{a}$ Deceased\\
% ...
$^{c}$ Also at the Novosibirsk State University, Novosibirsk, 630090, Russia\\
% ...
}
}

\begin{abstract}

Using a sample of $(2712\pm14)\times10^6$ $\psi(3686)$ events collected with the BESIII detector, we perform partial wave analyses of the decays $\psi(3686)\to p\bar{p}\pi^0$ and $\psi(3686)\to p\bar{p}\eta$. The branching fractions of $\psi(3686)\to p\bar{p}\pi^0$ and $\psi(3686)\to p\bar{p}\eta$ are determined to be $(133.9\pm11.2\pm2.3)\times10^{-6}$ or $(183.7\pm13.7\pm3.2)\times10^{-6}$ and $(61.5\pm6.5\pm1.1)\times10^{-6}$ or $(84.4\pm6.9\pm1.4)\times10^{-6}$, respectively, where the two solutions are caused by an ambiguous phase angle between resonant and continuum processes. Several well-established $N^*$ states are observed in the $p\pi^0$ and $p\eta$ systems, and the corresponding branching fractions are measured. The ratio of decay widths $\Gamma_{N(1535)\to N\eta}/\Gamma_{N(1535)\to N\pi}$ is determined to be $0.99\pm0.05\pm0.19$.
\end{abstract}

\maketitle

\section{Introduction}

The nature of the lowest lying $J^P=1/2^-$ nucleon resonance $N(1535)$ is not well understood as of yet. According to the conventional constituent quark model, the mass of the first orbital angular momentum excitation state, $N(1535)$, should always be below that of the first radial excitation state, $N(1440)$~\cite{Baryon_review}, also known as the Roper resonance. However, the experimental results indicate that the $N(1535)$ has a mass higher than $N(1440)$ by around 100 MeV, which has been a puzzle for several decades. Meanwhile, the measured partial width of  $N(1535)\to N\eta$ is much larger than expectation. An estimation based on flavor symmetry predicts a ratio of $\frac{\Gamma_{N(1535)\to N\eta}}{\Gamma_{N(1535)\to N\pi^0}}=0.17$~\cite{flavor_sym}. However, results from various multi-channel partial wave analyses~(PWAs) in fixed-target experiments listed by the Particle Data Group (PDG)~\cite{PDG} suggest an average value of $1.00\pm0.40$, which deviates from the prediction by 2$\sigma$ but suffers from poor precision.

Several theoretical studies have attempted to pin down the nature of $N(1535)$, explaining its mass puzzle and strong coupling to the $\eta$ meson~\cite{Baryon_review}. Some investigations~\cite{theory_1_1,theory_1_2,theory_1_3} suggest that $N(1535)$ is a dynamically generated quasi-bound state in the $K\Lambda$ and $K\Sigma$ channels, thus containing a sizable $s\bar{s}$ admixture. In some other investigations~\cite{theory_2_1,theory_2_2,theory_2_3}, the $N(1535)$ is assumed to have a penta-quark $uuds\bar{s}$ component. For these two scenarios, the assumption of five-quark contribution gives a natural explanation for the larger mass and enhanced coupling to $N\eta$ of $N(1535)$. Additionally, a study~\cite{theory_3_1} based on lattice QCD suggests a scenario where a three-quark core is dominating.

Experimentally, the decays of $\psi(3686)\to p\bar{p}\pi^0$ and $\psi(3686)\to p\bar{p}\eta$ have previously been studied by the CLEO and BESIII Collaborations. With $2.45\times10^7$ $\psi(3686)$ events, the CLEO collaboration reported evidence for several excited $N^*$ states in the $p\pi^0$ system in $\psi(3686)\to p\bar{p}\pi^0$ and in the $p\eta$ system in $\psi(3686)\to p\bar{p}\eta$ decays~\cite{exp_1}. Later, using $1.06\times10^8$ $\psi(3686)$ events collected in 2009, BESIII reported on PWAs of $\psi(3686)\to p\bar{p}\pi^0$~\cite{exp_2} and $\psi(3686)\to p\bar{p}\eta$~\cite{exp_3}, respectively. A ratio of decay widths $\frac{\Gamma_{N(1535)\to N\eta}}{\Gamma_{N(1535)\to N\pi^0}}=0.70\pm0.09^{+0.52}_{-0.32}$ was obtained with large systematic uncertainty, and no conclusion can be drawn from this measurement.  An improved measurement of this ratio is expected to strongly constrain the parameter space of the $N(1535)$, and hence to draw a firm conclusion on its deviation from flavor symmetry and pin down its nature.

The BESIII experiment has collected $(2712\pm14)\times10^6$ $\psi(3686)$ events, which provides an excellent opportunity to explore the nature of
the $N(1535)$. In this article, we present PWAs of
the decays $\psi(3686)\to p\bar{p}\pi^0$ and $\psi(3686)\to p\bar{p}\eta$ based on the full $\psi(3686)$ sample.

\section{DETECTOR AND DATA SAMPLES}
\label{sec:BES}

The BESIII detector~\cite{Ablikim:2009aa} records symmetric $e^+e^-$ collisions provided by the BEPCII storage ring~\cite{Yu:IPAC2016-TUYA01} in the center-of-mass energy ($\sqrt{s}$) range from 2.0 to  4.95~GeV,
with a peak luminosity of $1.1 \times 10^{33}\;\text{cm}^{-2}\text{s}^{-1}$ achieved at $\sqrt{s} = 3.77\;\text{GeV}$. BESIII has collected large data samples in this energy region~\cite{Ablikim:2019hff,EcmsMea,EventFilter}. The cylindrical core of the BESIII detector covers 93\% of the full solid angle and consists of a helium-based multilayer drift chamber~(MDC), a plastic scintillator time-of-flight system~(TOF), and a CsI(Tl) electromagnetic calorimeter~(EMC),
which are all enclosed in a superconducting solenoidal magnet providing a 1.0~T magnetic field. The solenoid is supported by an octagonal flux-return yoke with resistive plate counter muon
identification modules interleaved with steel.  The charged-particle momentum resolution at $1~{\rm GeV}/c$ is $0.5\%$, and the  specific energy loss  (${\rm d}E/{\rm d}x$) resolution is $6\%$ for electrons from Bhabha scattering. The EMC measures photon energies with a resolution of $2.5\%$ ($5\%$) at $1$~GeV in the barrel (end cap) region. The time resolution in the TOF barrel region is 68~ps, while that in the end cap region was 110~ps. The end cap TOF system was upgraded in 2015 using multigap resistive plate chamber technology, providing a time resolution of 60~ps~\cite{etof}. The data sample taken in 2021, about 83\% of the total data set used in this paper, benefits from this upgrade.

Monte Carlo (MC) simulated data samples produced with a {\sc geant4}-based~\cite{geant4} software package, which includes the geometric description of the BESIII detector and the detector response, are used to determine detection efficiencies and to estimate backgrounds. The simulation models the beam
energy spread and initial state radiation (ISR) in the $e^+e^-$ annihilations with the generator {\sc kkmc}~\cite{ref:kkmc}. The `inclusive' MC sample includes the production of the $\psi(3686)$ resonance, the ISR production of the $J/\psi$, and the continuum processes incorporated in {\sc kkmc}~\cite{ref:kkmc}. All particle decays are modeled with {\sc evtgen}~\cite{ref:evtgen} using branching fractions either taken from the PDG~\cite{PDG}, when available, or otherwise estimated with {\sc lundcharm}~\cite{ref:lundcharm}. To take into account the efficiency and luminosity difference between various rounds of data taking, both the exclusive and inclusive MC samples are simulated in different rounds separately, and then mixed together according to corresponding integrated luminosity.

\section{EVENT SELECTION AND BACKGROUND STUDY}
\label{sec:selection}

Candidates for $\psi(3686)\to p\bar{p}\pi^0,\pi^0\to\gamma\gamma$ and $\psi(3686)\to p\bar{p}\eta,\eta\to\gamma\gamma$ are required to have two charged tracks with opposite charge and at least two photon candidates. The charged tracks are required to satisfy $|V_z|< 10$~cm, $V_r< 1$~cm, and $\vert\!\cos\theta\vert< 0.93$. Here, $|V_z|$ and $V_r$ are the distances of the point of closest approach to the interaction point of the reconstructed track in the $z$ direction and the transverse ($x$-$y$) plane, respectively, and $\theta$ is the polar angle of charged track defined with respect to the $z$ axis.  The $z$ axis is defined as the symmetry axis of the MDC.  Photon candidates are reconstructed from isolated clusters in the
EMC. Their energies are required to be greater than 25 MeV in the
barrel ($\vert\!\cos\theta\vert<0.8$) and 50 MeV in the end cap
($0.86<\vert\!\cos\theta\vert<0.92$) regions. Clusters caused by
electronic noise or beam backgrounds are suppressed by requiring the
cluster time to be within [0, 700]~ns after the event start
time. To suppress fake photons produced by hadronic interactions in
the EMC and secondary photons from bremsstrahlung, clusters within a
cone of $20^\circ$ around the extrapolated position in the EMC
of any charged track are rejected.

To suppress the combinatorial background and improve the resolution, a five-constraint (5C) kinematic fit for the $\psi(3686)\to p\bar{p}\gamma\gamma$ hypothesis is performed, where in addition to overall energy-momentum conservation, the invariant mass of the two photons is constrained to the known mass of $\pi^0$ or $\eta$. The combination with the minimum $\chi^2_{\rm 5C}$ is kept for
further analysis. The $\chi_{\rm 5C}^{2}$ of the kinematic fit is
required to be smaller than 30.  The background from $J/\psi\to p\bar{p}$ is suppressed by rejecting events with $|M_{p\bar{p}}-3.097|<0.050$ GeV, where $M_{p\bar{p}}$ is the $p\bar{p}$ invariant mass.

\begin{figure*}[htbp]
	\centering
	\includegraphics[width=\textwidth]{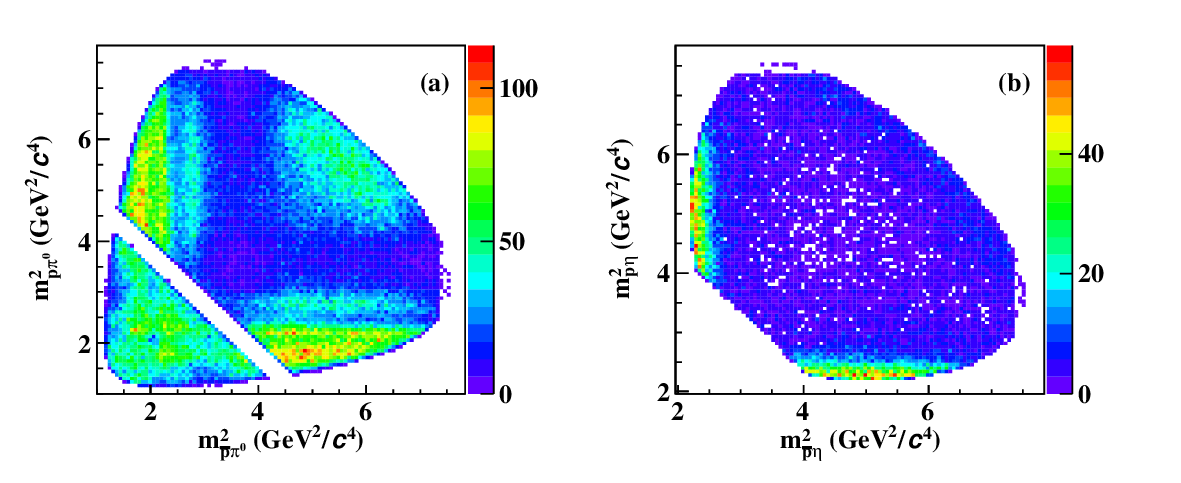}
	\caption{Dalitz plots of the selected (a) $\psi(3686)\to p\bar{p}\pi^0$ and (b) $\psi(3686)\to p\bar{p}\eta$ candidates from data.}
	\label{fig:Dalitz}
\end{figure*}

A total of 190,729 candidates for $\psi(3686)\to p\bar{p}\pi^0$ and 31,441 candidates for $\psi(3686)\to p\bar{p}\eta$ satisfy the above selection criteria. Dalitz plots are shown in Fig.~\ref{fig:Dalitz}, in which several excited $N^*$ states and $p\bar{p}$ structures can be seen. Studies of the $\psi(3686)$ inclusive MC sample indicate that the total background fraction $\frac{B}{S+B}$ from $\psi(3686)$ decays is 6.4\% for $\psi(3686)\to p\bar{p}\pi^0$, dominated by $\psi(3686)\to p\bar{p}\gamma^{\rm FSR}$ and $K/\pi$ misidentified as protons. For $\psi(3686)\to p\bar{p}\eta$, the background fraction is 9.5\%, and the dominant background contributions are from $\psi(3686)\to\gamma\chi_{cJ}$ with $\chi_{cJ}\to p\bar{p}\pi^0,p\bar{p}\gamma^{\rm FSR}$, where $\gamma^{\rm FSR}$ is the final state radiation photon.

To estimate the contribution from the continuum processes $e^+e^-\to p\bar{p}\pi^0$ and  $e^+e^-\to p\bar{p}\eta$, exactly the same selection criteria as above are applied to a data sample at $\sqrt{s}=3.773$~GeV with an integrated luminosity of $\mathcal{L}=2.93$~fb$^{-1}$. No evidence of $\psi(3770)\to p\bar{p}\pi^0$ is observed~\cite{exp_ee2pppi0} with current statistics; therefore this data sample is assumed to be dominated by the continuum process.

The interference between $\psi(3686)$ resonance and continuum process is studied by using the $\psi(3686)$ scan data samples collected from $\sqrt{s}=3.670$ to $3.710$ GeV with a total integrated luminosity $\mathcal{L}=417.7$~pb$^{-1}$. The charged tracks of the $p\bar{p}\eta$ final state are selected with the same criteria as above, while additional particle identification~(PID) by using  TOF and MDC ${\rm d}E/{\rm d}x$ information is imposed to the $p\bar{p}\pi^0$ final state to suppress peaking background. A charged track is identified as a proton if its PID confidence level for the proton hypothesis is larger than that for the pion and kaon hypotheses. The selection for photon candidates is kept to be the same as for the $\psi(3686)$ data sample.  A four-momentum constraint (4C) kinematic fit is performed  and the corresponding $\chi^2_{\rm 4C}$ is required to be less than 30. The veto on the $J/\psi\to p\bar{p}$ background is also applied.

\section{PARTIAL WAVE ANALYSIS}
\label{sec:PWA}

\subsection{Amplitude}

In the PWA, the decay amplitudes for the sequential decays, via an intermediate states of a neutral meson $M^0$ or charged $N^{*\pm}$,
\begin{equation}
	\begin{aligned}
	&\psi(3686)\to M^0 \pi^0(\eta), M^0\to p\bar{p},\\
	&\psi(3686)\to N^{*+} \bar{p},N^{*+}\to p\pi^0(\eta),\\
	&\psi(3686)\to \bar{N}^{*-} p,\bar{N}^{*-}\to \bar{p}\pi^0(\eta),\end{aligned}
\label{eq:decay_chain}
\end{equation}
are constructed using the relativistic covariant tensor amplitude formalism~\cite{FDC}.
The structure for the baryon-antibaryon-meson vertices is composed by $p_i^\mu,~\gamma_\mu,~g^{\mu\nu},~\epsilon^{\alpha\beta\rho\sigma}$ and their products or contractions, where $p_i$ are the momenta of the particles involved. These effective vertices are deduced from an effective Lagrangian
\begin{equation}
\mathcal{L} = \bar\psi_1\Gamma \psi_2 A,
\end{equation}
where $\Gamma$ is the effective vertex, and $\psi_{1,2}$ and $A$ are the baryon and virtual photon fields, respectively. Here, the Lorentz indices are omitted. 
For the strong decays considered in this work (Eq.~\ref{eq:decay_chain}), the corresponding interaction Lagrangian must be C- and P-parity invariant, Lorentz invariant, and CPT invariant. These constraints lead to the strong interaction vertices satisfying
\begin{equation}
\Gamma = \xi_AC(\gamma_0\Gamma^+\gamma_0)^TC^{-1}\text{ and } \Gamma = \eta_1\eta_2\eta_A\gamma_0\Gamma^P\gamma_0,
\end{equation}
where $\xi_A$ is the C parity of the virtual photon; $C=-i\gamma_2\gamma_0$ is the C transformation operator; $\Gamma^+$ is the Hermitian conjugation of $\Gamma$; $\eta_A$ and $\eta_{1,2}$ are the P parities of meson and baryons, respectively; $\Gamma^P$ is the vertex $\Gamma$ after P-parity transformation.

The amplitude $\mathcal{A}_j$ of a decay via a resonance is
\begin{equation}
    \mathcal{A}_j=\epsilon^{*\alpha}(p_0,m)\bar u(p_1,s_1)\Gamma_{1\alpha\mu_1\mu_2\ldots}\Gamma_2^{\mu_1\mu_2\ldots}v(p_2,s_2){\rm BW}(s),
\end{equation}
where $\epsilon^{*}$ is the $\psi(3686)$ polarization vector; $u(p_1,s_1)$ and $v(p_2,s_2)$ are the free Dirac spinors for proton and anti-proton, respectively; $\Gamma_1$ and $\Gamma_2$ are the strong interaction vertices for the resonances with $\psi(3686)$ and $p$, $\bar{p}$ and $\pi^0$ or $\eta$; ${\rm BW}(s)$ is a Breit-Wigner function.

The decay dynamics of the three lowest lying $N^*$ states are described with the KSU model~\cite{KSU_model}. The energy-dependent Breit-Wigner width is parametrized as
\begin{equation}
	\Gamma(\sqrt{s^\prime})=\Gamma_0\times\sum_{i}r_{i}\times\frac{\rho_i(\sqrt{s^\prime})}{\rho_i(m_0)},
\end{equation}
where $\sqrt{s^\prime}$ is the two-body invariant mass of $p\pi^0$ or $p\eta$, $m_0$ is the mass of $N^*$, and $r_i$ is the branching fraction of the $i$-th decay channel of $N^*$ as summarized in Table~\ref{tab:Rparas}.
 \begin{table}[htbp]
	\centering
	\tabcolsep=0.16cm
	\small
	\caption{The dominant decay channels and corresponding branching fractions of the three lowest lying $N^*$ in unit of percent~\cite{KSU_model}.}
	\begin{tabular}{c|c|c|c|c|c|c}
		\hline
		\hline
		Resonance &\multicolumn{2}{c|}{$N(1440)$} & \multicolumn{2}{c|}{$N(1520)$} & \multicolumn{2}{c}{$N(1535)$} \\
		\hline
		\multirow{4}{*}{Decay} & $N\pi$& $60\pm2$ & $N\pi$ &$59\pm2$ & $N\pi$ & $43\pm2$\\
		&$\Delta\pi$ & $23\pm4$  &$\Delta\pi$(S-wave) & $21\pm2$ & $N\eta$ & $43\pm2$ \\
		&$N\sigma$ & $17\pm3$  &$\Delta\pi$(D-wave) & $6\pm1$ &$N\rho$ & $14\pm2$ \\
		&--- & --- & $N\rho$ & $14\pm2$ &--- & --- \\
		\hline
		\hline
	\end{tabular}
	\label{tab:Rparas}
\end{table}
For the two-body decay of a resonance into the $i$-th channel with two stable particles, the phase-space factor is parametrized as
\begin{equation}
	\rho_i = \frac{q_i}{\sqrt{s^\prime}}\times B^2_l(q_iR),
\end{equation}
where $q_i$ is the momentum of the two particles in their center-of-mass frame, $B_{l_i}$ is the Blatt-Weisskopf barrier penetration factor~\cite{Barrier_factor}, and $l_i$ is the orbital angular momentum of the two particles. The range parameter $R$ is fixed at 0.73 fm. The first three Blatt-Weisskopf factors are given as
\begin{equation}
	\begin{split}
		B^2_0(x) &= 1,\\
		B^2_1(x) &= \frac{x^2}{1+x^2},\\
		B^2_2(x) &= \frac{x^4}{9+3x^2+x^4},
	\end{split}
\end{equation}
where $x=q_iR$. For a quasi-two-body decay of a resonance into the $i$-th channel consisting of a stable particle of mass $m$ and an isobar of mass $M$, which is assumed to decay into stable particles with masses $m_1$ and $m_2$, the phase-space factor is calculated numerically by
\begin{equation}
    \rho_i=\int^{\sqrt{s^\prime}-m}_{m_1+m_2}\frac{q}{\sqrt{s^\prime}}\times B^2_l(qR)\times {\rm BW}^\prime(M) \times {\rm d}M.
\end{equation}
Here ${\rm BW}^\prime(M)$ is the line shape of the isobar, which is parametrized by a Breit-Wigner function with a constant width, E791 type~\cite{Sigma500_lineshape}, and the Gounaris-Sakurai model~\cite{Rho770_lineshape} for $\Delta$, $\sigma$, and $\rho$, respectively.  For the other intermediate states, the constant width $\Gamma(m)=\Gamma_{0}$ is used. In the fit, all the masses and widths are fixed to the central values shown in Table~\ref{tab:baseline}.

\subsection{Fit method}
The complex coupling constants of the amplitudes are determined by an unbinned maximum likelihood fit. The probability to observe the $i$-th event characterized by $\xi_i$, \emph{i.e.}, the measured four momenta of the particles in the final state, is
\begin{equation}
	p(\xi_i)=\frac{\omega(\xi_i)\epsilon(\xi_i)}{\int d\Phi \omega(\xi)\epsilon(\xi)},
\end{equation}
where $\omega(\xi_i)\equiv(\frac{d\sigma}{d\Phi})_i$ is the differential cross section, $\epsilon$ is the detection efficiency, $d\Phi$ is the standard element of phase space for the three-body decays and $\sigma^\prime = \int d\Phi \omega(\xi)\epsilon(\xi)$ is the observed total cross section. The differential cross section is
\begin{equation}
	\omega(\xi)\equiv\frac{d\sigma}{d\Phi}=\left|\sum_j\mathcal{A}_j\right|^2,
\end{equation}
where $A_j$ is the amplitude for the $j$-th resonance.

The likelihood for the data sample is taken as
\begin{equation}
	\mathcal{L}=\prod_{i=1}^{N}P(\xi_i)=\prod_{i=1}^{N}\frac{\omega(\xi_i)\epsilon(\xi_i)}{\sigma^\prime}.
\end{equation}
Technically, the {\sc minuit} package~\cite{Minuit} is used to minimize a negative log-likelihood~(NLL) function, $\mathcal{S}=-\ln\mathcal{L}$, instead of maximizing $\mathcal{L}$, with
\begin{equation}
	\mathcal{S}=-\ln \mathcal{L}=-\sum_i^N\ln\left[\frac{\omega(\xi_i)}{\sigma^\prime}\right]-\sum_{i}^{N}\ln\epsilon(\xi_i).
	\label{eq:pwa_likelihood_function}
\end{equation}
In Eq.~\ref{eq:pwa_likelihood_function}, the second term is independent on the parameters and has no impact on the determination of the parameters or on the relative changes in $\mathcal{S}$. In the fit, $-\ln\mathcal{L}$ is reduced to
\begin{equation}
	\mathcal{S}=-\sum_{i}^{N}\ln\omega(\xi_i)+N\ln\sigma^\prime.
\end{equation}

The observed total cross section $\sigma^\prime$ is evaluated using an MC sample consisting of $N_{\rm gen}$ signal events distributed uniformly in phase space. These events are subjected to the selection criteria described in Sec.~\ref{sec:selection} and yield a sample of $N_{\rm acc}$ accepted events. The normalization integral is then computed as
\begin{equation}
	\int \mathrm{d} \Phi \omega(\xi) \epsilon(\xi)=\sigma^{\prime} = \frac{1}{N_{\rm{gen}}} \sum_{k}^{N_{\rm{acc}}} \omega(\xi_{k}).
\end{equation}

The background contribution in the fit is estimated using the inclusive MC sample and is subtracted from the $-\ln\mathcal{L}$ function by
\begin{equation}
	\mathcal{S}=-(\ln\mathcal{L}_{\rm data}-\ln\mathcal{L}_{\rm bkg}).
\end{equation}

After the parameters are determined in the fit, the signal yield of a given resonance is estimated by scaling its cross section ratio $r_j$ to the net number of events
\begin{equation}\label{yieldsFormula}
N_j=r_j(N_\textrm{\scriptsize{data}}-N_\textrm{\scriptsize{bkg}}),\textrm{~with~}r_j={\sigma^\prime_j\over \sigma^\prime},
\end{equation}
where $\sigma^\prime_j=\int d\Phi \omega_j(\xi) \epsilon(\xi)$ is the observed cross section for the $j$-th resonance with $\omega_j=|\mathcal{A}_j|^2$, $N_\textrm{data}$ is the number of observed events in data and $N_\textrm{bkg}$ is the number of background events.

The statistical uncertainty of the signal yield $N_i$, $\delta N_{i}$, is estimated according to the uncertainty propagation formula using the covariance matrix obtained in the fit.

\subsection{Amplitude analysis results of $\psi(3686)\to p\bar{p}\pi^0$}~\label{sec:pi0_pwa}

A PWA is performed to the selected 190,729 candidate events for $\psi(3686)\to p\bar{p}\pi^0$, where the background yield is determined to be 12,172 with the inclusive MC sample.

The $N^*$ states with $J<\frac{5}{2}$ and certain existence of $N^*\to N\pi$ from the PDG~\cite{PDG}, {\it i.e.} $N(1440)$, $N(1520)$, $N(1535)$, $N(1710)$, and $N(1720)$, are considered in the significance test. Additionally, the $N(2100)$, $N(2300)$, and $N(2570)$ are included according to the previous studies on $J/\psi\to p\bar{n}\pi^-+c.c.$~\cite{exp_n2100} and $\psi(3686)\to p\bar{p}\pi^0$~\cite{exp_2}. 
According to the framework of soft $\pi$ meson theory~\cite{ref::th_N940}, an off shell process $N(940)\to p\pi^0$ is included. The significant contribution from the virtual proton pole $N(940)$ is expected~\cite{ref::th_N940_1} and also observed by several experiment studies~\cite{exp_n2100,ref::Jpsi_pppi0,exp_2}. Thus, $N(940)$ is also included into significance test.
The $p\bar{p}$ structures are explained with $\rho(1900)$~\cite{exp_rho1900}, $\rho(2000)$~\cite{exp_rho2000}, $\rho(2150)$~\cite{exp_rho2150}, and $\rho(2225)$~\cite{exp_rho2225}. The statistical significance of those intermediate states are estimated by removing one of them from solution at one time. Based on changes in the log-likelihood and the number of degrees of freedom, the statistical significance is calculated. All intermediate states yield significance greater than 5$\sigma$, thus are kept in the baseline solution. Table~\ref{tab:baseline} shows the resonance parameters and quantum numbers of those intermediate states.

 \begin{table*}[htbp]
	\centering
	\caption{The resonance parameters, quantum numbers, and references of intermediate states involved in the baseline solutions $\psi(3686)\to p\bar{p}\pi^0$ and $\psi(3686)\to p\bar{p}\eta$.}
	\begin{tabular}{c|cccc}
		\hline
		\hline
		Resonance state & Mass (MeV/$c^2$) & Width (MeV) & $I^G(J^{PC})$ & comment \\
		\hline
		$\rho(1900)$  &  $1880\pm30$  & $130\pm30$ & $1^+(1^{--})$ & Ref.~\cite{exp_rho1900} \\
		\hline
		$\rho(2000)$  &  $2078\pm6$  & $149\pm21$ & $1^+(1^{--})$ & Ref.~\cite{exp_rho2000}\\
		\hline
		$\rho(2150)$  &  $2254\pm22$  & $109\pm76$ & $1^+(1^{--})$ & Ref.~\cite{exp_rho2150}\\
		\hline
		$\rho(2225)$  &  $2225\pm35$  & $335\pm100$ & $1^+(2^{--})$ & Ref.~\cite{exp_rho2225}\\
		\hline
		$\phi_3(1850)$  &  $1854\pm7$  & $87\pm28$ & $0^-(3^{--})$ & PDG~\cite{PDG} \\
		\hline
		$\omega(1960)$  &  $1960\pm25$  & $195\pm60$ & $0^-(1^{--})$ & Ref.~\cite{exp_omega1960}\\
		\hline
		$\omega(2205)$  &  $2205\pm30$  & $350\pm90$ & $0^-(1^{--})$ & Ref.~\cite{exp_omega1960} \\
		\hline
		$N(940)$  & 940  & 0 & $\frac{1}{2}(\frac{1}{2}^+)$ & Virtual proton pole \\
		\hline
		$N(1440)$&  $1406\pm3$& $314\pm9$  & $\frac{1}{2}(\frac{1}{2}^+)$ & Optimized with data\\
		\hline
		$N(1520)$ &  $1512\pm2$& $121\pm3$  & $\frac{1}{2}(\frac{3}{2}^-)$ & Ref.~\cite{KSU_model}\\
		\hline
		$N(1535)$ & $1525\pm2$ & $147\pm5$  & $\frac{1}{2}(\frac{1}{2}^-)$ & Ref.~\cite{KSU_model}\\
		\hline
		$N(1650)$ & $1666\pm3$  & $133\pm7$  & $\frac{1}{2}(\frac{1}{2}^-)$ & Ref.~\cite{KSU_model}\\
		\hline
		$N(1710)$ & $1710\pm30$ & $140\pm60$  & $\frac{1}{2}(\frac{1}{2}^+)$ & PDG~\cite{PDG} \\
		\hline
		$N(1720)$ & $1720\pm30$ & $250\pm100$  & $\frac{1}{2}(\frac{3}{2}^+)$ & PDG~\cite{PDG} \\
		\hline
		$N(1895)$ & $1895\pm25$  & $120\pm80$  & $\frac{1}{2}(\frac{1}{2}^-)$ & PDG~\cite{PDG} \\
		\hline
		$N(2100)$ & $2100\pm50$  & $260\pm60$  & $\frac{1}{2}(\frac{1}{2}^+)$ & PDG~\cite{PDG} \\
		\hline
		$N(2300)$ & $2300\pm116$  & $340\pm114$  & $\frac{1}{2}(\frac{1}{2}^+)$ & PDG~\cite{PDG}\\
		\hline
		$N(2570)$ & $2570\pm39$  & $250\pm70$  & $\frac{1}{2}(\frac{5}{2}^-)$ & PDG~\cite{PDG}\\
		\hline
		\hline
	\end{tabular}
	\label{tab:baseline}
\end{table*}

\begin{figure*}[htbp]
	\centering
	\includegraphics[width=\textwidth]{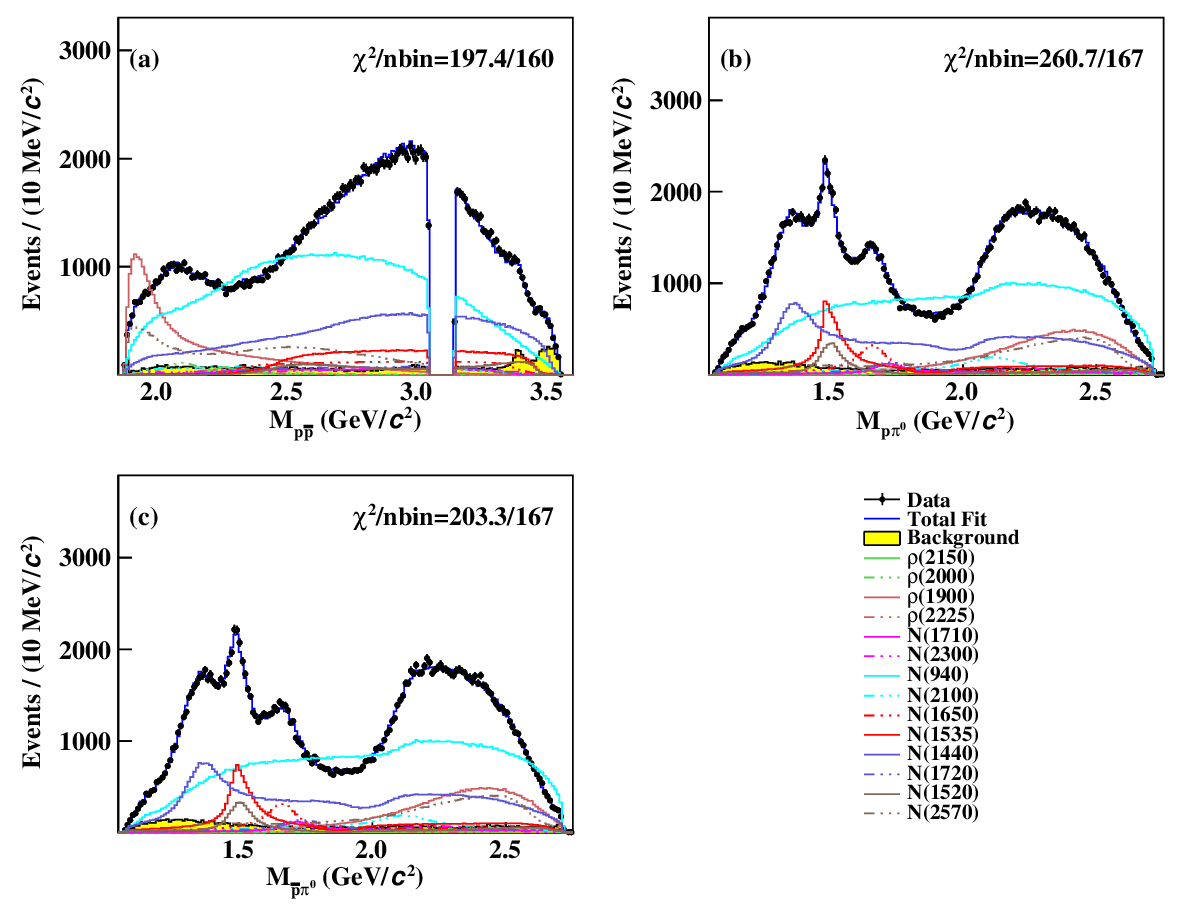}
	\caption{Projections of the PWA fit at $\sqrt{s}=3.686$ GeV on the (a) $p\bar{p}$, (b) $p\pi^0$, and (c) $\bar{p}\pi^0$ invariant mass distributions. The points with errors are data, the blue curves are the fit result, the curves in various colors denote different resonant components and the yellow filled histograms are the simulated background.}
	\label{fig:PWA_fit_pi0_mass}
\end{figure*}

Figure~\ref{fig:PWA_fit_pi0_mass} shows the projections of PWA fit result on the two-body invariant mass spectra $M_{p\pi^0}$, $M_{\bar{p}\pi^0}$, and $M_{p\bar{p}}$, respectively. The fit result gives a good description of the data sample. The fitted signal yields of $N^*$ states at $\sqrt{s}=3.686$ GeV, denoted as $N_{\rm sig}$, are summarized in Table~\ref{tab:pi0_result_pwa}.

\begin{figure*}[htbp]
	\centering
	\includegraphics[width=\textwidth]{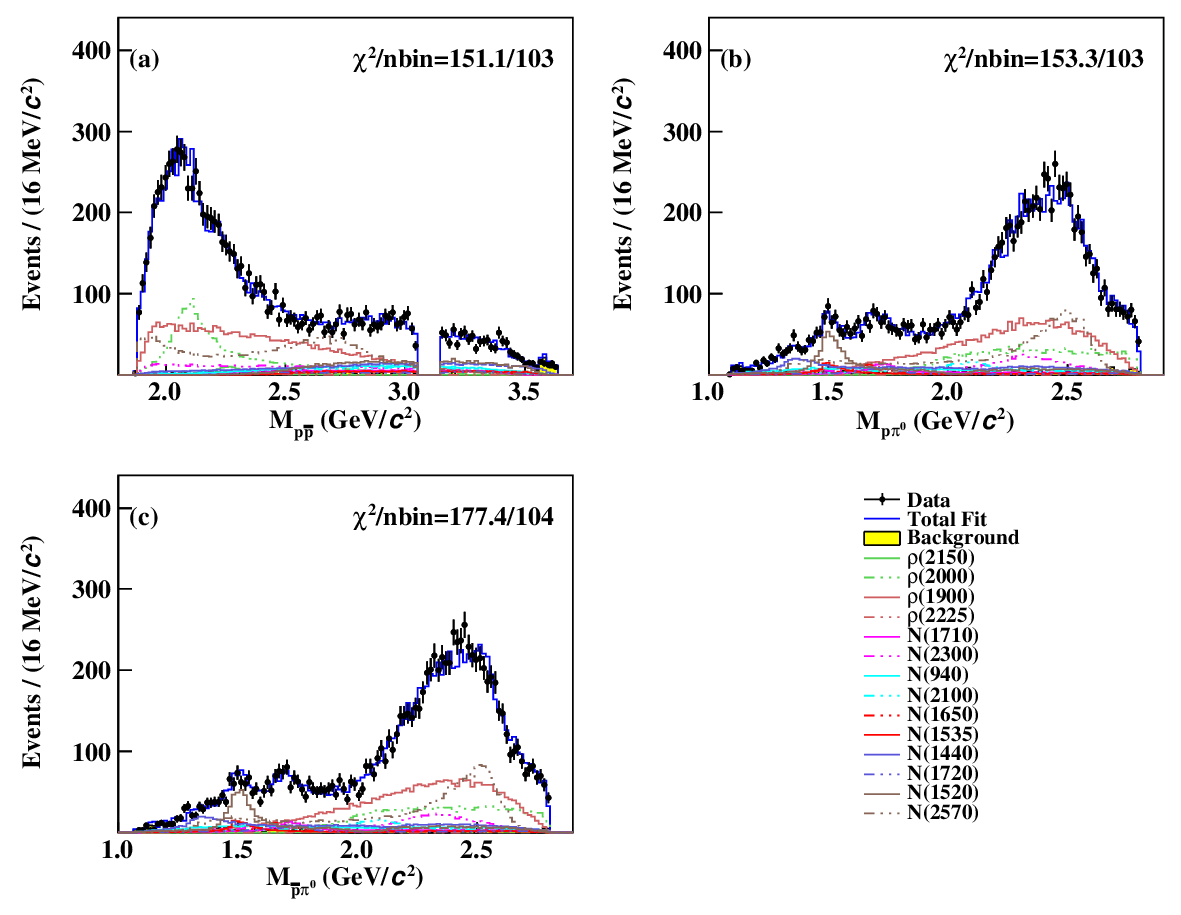}
	\caption{Projections of the PWA fit at $\sqrt{s}=3.773$ GeV on the (a) $p\bar{p}$, (b) $p\pi^0$, and (c) $\bar{p}\pi^0$ invariant mass distributions. The points with errors are data, the blue curves are the fit result, the curves in various colors denote different resonant components and the yellow filled histograms are the simulated background.}
	\label{fig:PWA_fit_pi0_mass_3773}
\end{figure*}

The contribution from the continuum process is studied by performing the same PWA fit to the data sample at $\sqrt{s}=3.773$~GeV, as shown in Fig.~\ref{fig:PWA_fit_pi0_mass_3773}. The fitted signal yields of $N^*$ states at $\sqrt{s}=3.773$~GeV, denoted as $N_{\rm con}$, are also summarized in Table~\ref{tab:pi0_result_pwa}. Potential contributions from the isospin violating processes $\gamma^*\to (\omega/\phi)^*\pi^0$ are tested and found to be insignificant. Considering the significant contribution of $\rho^*$ states from continuum process and potential strong interference between $\psi(3686)$ resonance and continuum production, only the branching fractions of $N^*$ states are reported. The yield $N_{\rm con}$ is further rescaled to $\sqrt{s}=3.686$ GeV with the factor
\begin{equation}
\begin{split}
f_c=\frac{\mathcal{L}^{3686}\times\sigma_{\rm Born }^{3686}\times(\epsilon(1+\delta))^{3686}}{\mathcal{L}^{3773}\times\sigma_{\rm Born}^{3773}\times(\epsilon(1+\delta))^{3773}}
\end{split}
\label{eq:normalization_factor}
\end{equation}
by considering the integrated luminosity $\mathcal{L}$, Born cross section $\sigma_{\rm Born}$, efficiency $\epsilon$, and initial state radiation~(ISR) correction $(1+\delta)$. The integrated luminosities at $\sqrt{s}=3.686$~GeV and $\sqrt{s}=3.773$~GeV are 4.08~fb$^{-1}$ and 2.93~fb$^{-1}$, respectively. The Born cross sections of $e^+e^-\to p\bar{p}\pi^0$ are estimated with Ref.~\cite{exp_ee2pppi0}. The efficiency is obtained by generating MC samples with the PWA fit result at $\sqrt{s}=3.773$~GeV. The radiative correction factor $(1+\delta)$ is given by {\sc conexc}~\cite{conexc} based on the Born cross section. The $f_c$ is determined to be $1.564$.

The net signal yield is obtained by subtracting the normalized contribution from the continuum process, denoted as $N_{\rm net}$, and is calculated with $N_{\rm net}=N_{\rm sig}-f_c\times N_{\rm con}$. The signal efficiency~$\epsilon$ for each intermediate state is obtained from a generated corresponding signal MC sample according to the PWA fit result at $\sqrt{s}=3.686$ GeV. The branching fractions are given by
\begin{equation}
\begin{split}
\mathcal{B}=\frac{N_{\rm net}}{\epsilon\times\mathcal{B}(\pi^{0} \to\gamma\gamma)\times N_{\psi(3686)}},
\end{split}
\end{equation}
as summarized in Table~\ref{tab:pi0_result_pwa}.  Compared to the previous study on $\psi(3686)\to p\bar{p}\pi^0$~\cite{exp_2}, there are some mild tensions, which are attributed mainly to differences in the treatment of continuum backgrounds in the PWA and several newly-included $\rho^*$ and $N^*$ states in the baseline solution.


\begin{table*}
	\centering
	\tabcolsep=0.5cm
	\caption{Numbers used to calculate branching fractions of $N^*$ states in $\psi(3686)\to p\bar{p}\pi^0$. $N_{\rm sig}$ is the fitted yield from PWA of $\psi(3686)$ data sample. $N_{\rm con}$ is the fitted yield from PWA of continuum data sample. $N_{\rm net}$ is the net signal yield calculated with $N_{\rm net}=N_{\rm sig}-f_c\times N_{\rm con}$, where $f_c$ is the normalization factor, $\epsilon$ is signal efficiency and $\mathcal{B}$ is branching fraction. Here, first uncertainties are statistical and the second are systematic.}
	\begin{tabular}{cccccc}
		\hline
		\hline Resonance state & $N_{\rm sig}$ & $N_{\rm con}$ & $N_{\rm net}$ & $\epsilon$ (\%) & $\mathcal{B}$ ($\times10^{-6}$) \\
		\hline $N(940)$ & $122215\pm3266$  & $656\pm164$ & $121188\pm3276$ & $39.71$ & $113.9\pm3.1\pm11.9$ \\
		\hline$N(1440)$ & $57118\pm1383$  & $953\pm147$ & $55627\pm1402$ & $38.34$ & $54.2\pm1.4\pm14.1$ \\
		\hline$N(1520)$ & $8109\pm428$  & $870\pm81$ & $6749\pm446$ & $38.43$ & $6.6\pm0.4\pm2.0$ \\
		\hline$N(1535)$ & $18894\pm778$  & $240\pm77$ & $18519\pm787$ & $39.61$ & $17.5\pm0.7\pm3.6$ \\
		\hline$N(1650)$ & $11146\pm794$  & $278\pm79$ & $10712\pm804$ & $43.75$ & $9.1\pm0.7\pm3.2$ \\
		\hline$N(1710)$ & $5043\pm472$  & $369\pm100$ & $4466\pm497$ & $39.73$ & $4.2\pm0.5\pm4.0$ \\
		\hline$N(1720)$ & $6983\pm523$  & $217\pm83$ & $6644\pm539$ & $39.93$ & $6.2\pm0.5\pm2.4$ \\
		\hline$N(2100)$ & $11107\pm1033$  & $551\pm161$ & $10245\pm1063$ & $44.90$ & $8.5\pm0.9\pm3.8$ \\
		\hline$N(2300)$ & $5633\pm566$  & $894\pm222$ & $4235\pm664$ & $43.75$ & $3.6\pm0.6\pm3.0$ \\
		\hline$N(2570)$ & $27716\pm1041$  & $2349\pm187$ & $24043\pm1082$ & $46.14$ & $19.5\pm0.9\pm14.5$ \\
		\hline
		\hline
	\end{tabular}
	\label{tab:pi0_result_pwa}
\end{table*}

\subsection{Amplitude analysis results of $\psi(3686)\to p\bar{p}\eta$}

A PWA is performed to the selected 31,441 candidate events for $\psi(3686)\to p\bar{p}\eta$, where the background yield is determined to be 2,986 with the inclusive MC sample.

The $N^*$ states with at least fair existence of $N^*\to N\eta$ from the PDG~\cite{PDG}, {\it i.e.} $N(1535)$, $N(1650)$, $N(1710)$, and $N(1895)$, are considered in the significance test. The $N(1520)$ is not included since the $\Gamma_{N\eta}/\Gamma_{\rm tot}$ is smaller than 1\% for this resonance. The $p\bar{p}$ structures are explained with several excited $\omega$ and $\phi$ states. The statistical significance of all $N^*$ states and some of the structures in $p\bar{p}$,  $\phi_3(1850)$, $\omega(1960)$~\cite{exp_omega1960}, and $\omega(2205)$~\cite{exp_omega1960} are greater than 5$\sigma$, hence they are kept in the baseline solution. The contribution from the virtual proton pole $N(940)$ in the $p\bar{p}\eta$ final state is expected to be negligible~\cite{ref::th_N940} and also found to be insignificant in the PWA fit.  Table~\ref{tab:baseline} shows the resonance parameters and quantum numbers of those intermediate states.

\begin{figure*}[htbp]
	\centering
	\includegraphics[width=\textwidth]{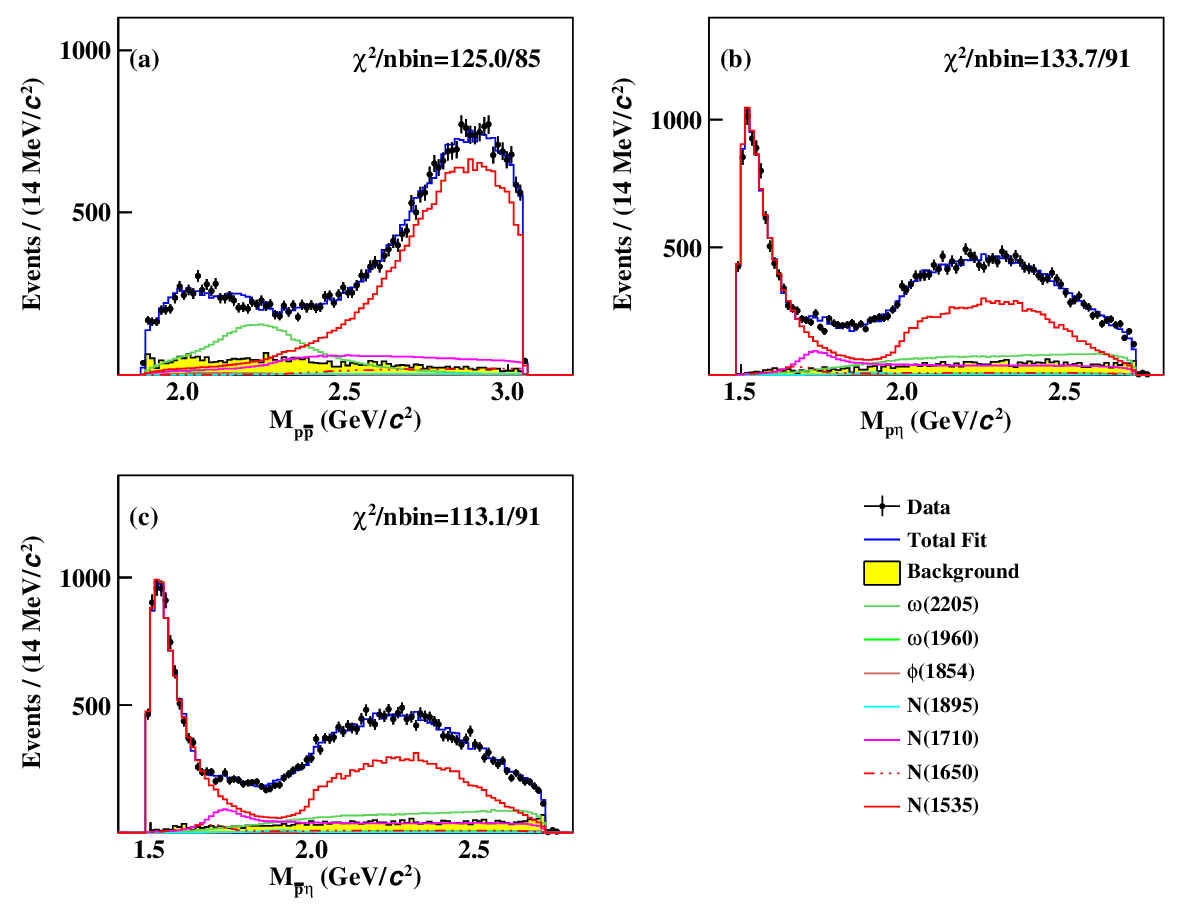}
	\caption{Projections of the PWA fit at $\sqrt{s}=3.686$ GeV on the (a) $p\bar{p}$, (b) $p\eta$, and (c) $\bar{p}\eta$ invariant mass distributions. The points with errors are data, the blue curves are the fit result, the curves in various colors denote different resonant components and the yellow filled histograms are the simulated background.}
	\label{fig:PWA_fit_eta_mass}
\end{figure*}

Figure~\ref{fig:PWA_fit_eta_mass} shows the projection of  PWA result on the two-body invariant mass spectra $M_{p\eta}$, $M_{\bar{p}\eta}$, and $M_{p\bar{p}}$, respectively. The distribution of data sample is well described by the fit result.

\begin{figure*}[htbp]
	\centering
	\includegraphics[width=\textwidth]{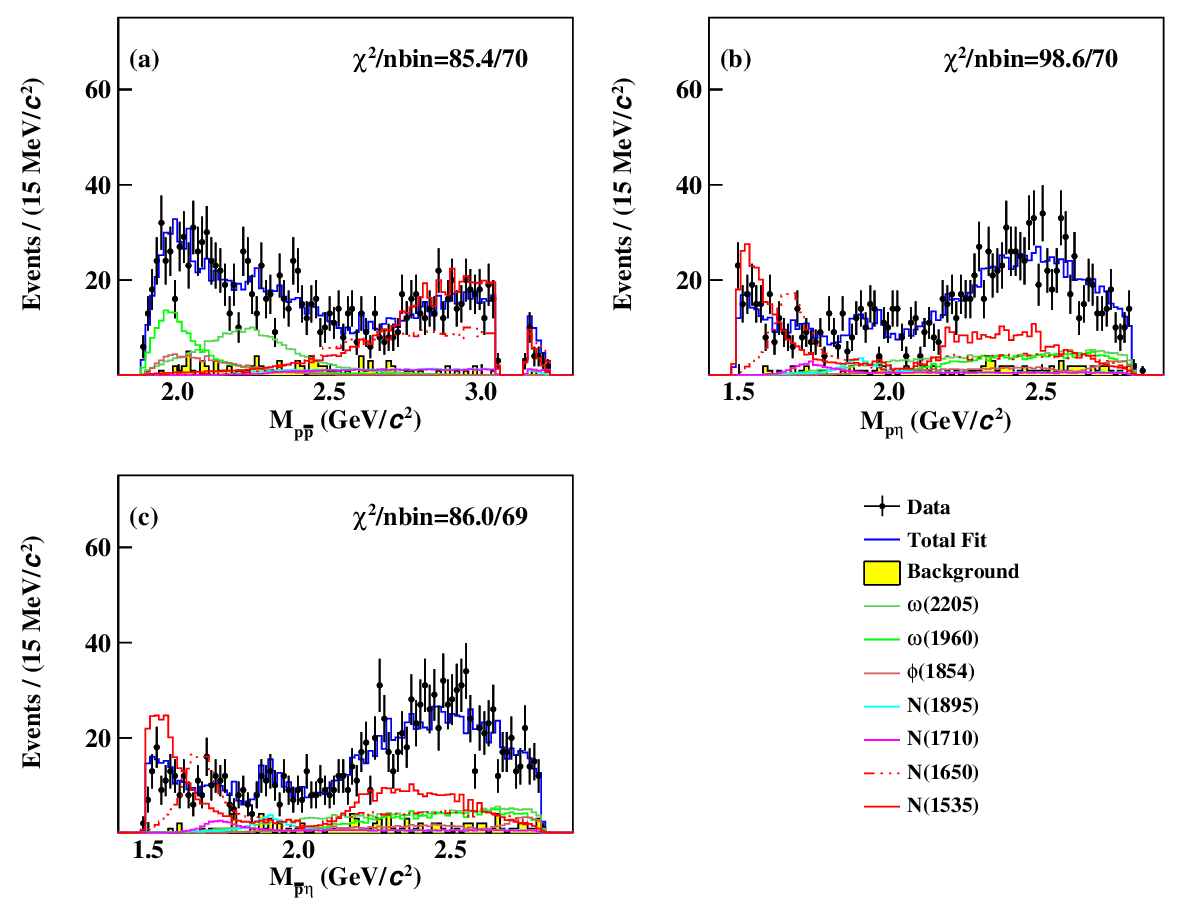}
	\caption{Projections of the PWA fit at $\sqrt{s}=3.773$ GeV on the (a) $p\bar{p}$, (b) $p\eta$, and (c) $\bar{p}\eta$ invariant mass distributions. The points with errors are data, the blue curves are the fit result, the curves in various colors denote different resonant components and the yellow filled histograms are the simulated background.}
	\label{fig:PWA_fit_eta_mass_3773}
\end{figure*}

The same study as described above in Sec.~\ref{sec:pi0_pwa} to estimate the continuum contribution is performed.  The results are shown in Fig.~\ref{fig:PWA_fit_eta_mass_3773}. Based on the Born cross sections of $e^+e^-\to p\bar{p}\eta$ taken from Ref.~\cite{exp_ee2ppeta}, the normalization factor $f_c$ is determined to be $1.623$ according to Eq.~\ref{eq:normalization_factor}. The parameters $N_{\rm sig}$, $N_{\rm con}$, $N_{\rm net}$, and $\epsilon$ for $\psi(3686)\to p\bar{p}\eta$ have exactly the same definition as in Sec.~\ref{sec:pi0_pwa}. The branching fractions are given with
\begin{equation}
\begin{split}
\mathcal{B}=\frac{N_{\rm net}}{\epsilon\times\mathcal{B}(\eta \to\gamma\gamma)\times N_{\psi(3686)}},
\end{split}
\end{equation}
and summarized in Table~\ref{tab:eta_result_pwa}. For $N(1650)$ and $N(1895)$, the $N_{\rm net}$ value are consistent with zero within two times the statistical uncertainties;  the corresponding branching fractions are not reported. The determined branching fraction of $N(1535)$ is in good agreement with the previous measurement~\cite{exp_3}.

\begin{table*}[htbp]
	\centering
	\tabcolsep=0.5cm
	\caption{Numbers used to calculate branching fractions of $N^*$ states in $\psi(3686)\to p\bar{p}\eta$. The definition is exactly the same as that of Table~\ref{tab:pi0_result_pwa}. Here, the first uncertainties are statistical and the second are systematic.
	}
	\begin{tabular}{cccccc}
		\hline
		\hline Resonance state & $N_{\rm sig}$ & $N_{\rm con}$ & $N_{\rm net}$ & $\epsilon$ (\%) & $\mathcal{B}$ ($\times10^{-6}$) \\
		\hline$N(1535)$ & $20411\pm460$ & $570\pm115$ & $19486\pm486$ & $36.17$ & $50.5\pm1.3\pm7.1$ \\
		\hline$N(1650)$ & $809\pm310$ & $388\pm88$ & $180\pm341$ & --- & --- \\
		\hline$N(1710)$ & $3351\pm273$ & $63\pm63$ & $3250\pm292$ & $38.81$ & $7.8\pm0.7\pm3.1$ \\
		\hline$N(1895)$ & $198\pm50$ & $71\pm32$ & $83\pm72$ & --- & --- \\
		\hline
		\hline
	\end{tabular}
	\label{tab:eta_result_pwa}
\end{table*}

\section{OBSERVED CROSS SECTION}

The branching fractions of $\psi(3686)\to p\bar{p}\pi^0$ and $\psi(3686)\to p\bar{p}\eta$ is determined by fitting the observed cross sections at nine energy points around the $\psi(3686)$ peak, including both the $\psi(3686)$ data sample and the $\psi(3686)$-scan data sample. Here, the $\psi(3686)$ data sample is selected in the same way as that of scan data sample.

\begin{figure*}[htbp]
	\centering
	\includegraphics[width=\textwidth]{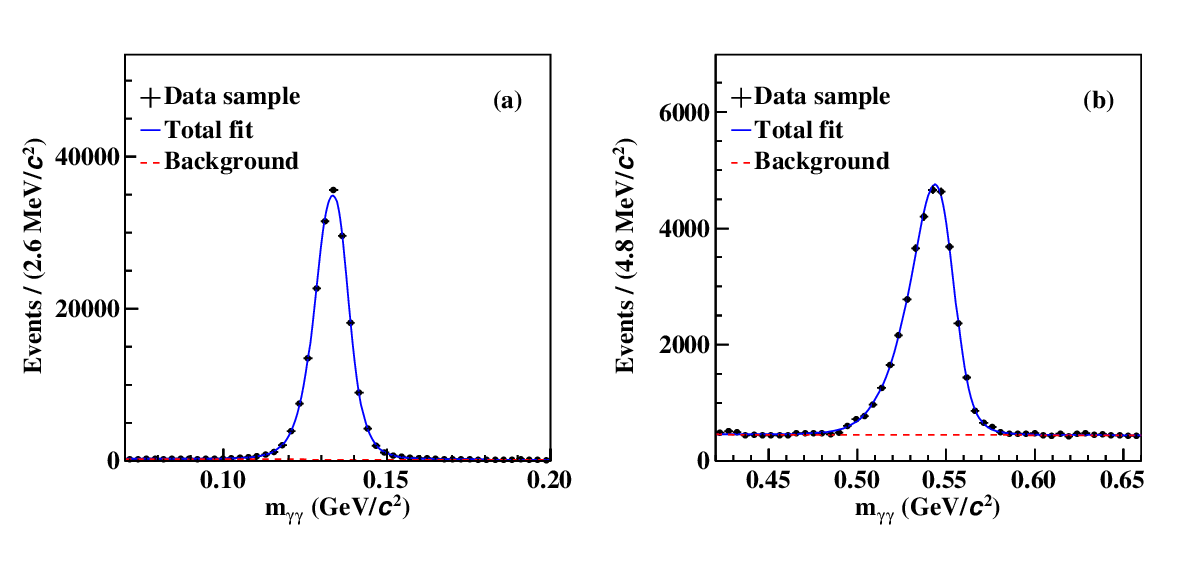}
	\caption{Fits to the $m_{\gamma\gamma}$ spectra of the selected candidates of the $\psi(3686)$ data sample in (a) $\pi^0$ and (b) $\eta$ signal zones. The points with errors are the $\psi(3686)$ data, the blue solid curves are the fit results, and the red dashed curves are the background contributions.}
	\label{fig:fit_at_3686}
\end{figure*}

The signal yield $N_{\rm obs}$ at each energy point is determined by a fit to the $m_{\gamma\gamma}$ spectrum. The signal shape is described with the MC simulated shape convolved with a Gaussian function to take into account the resolution difference between data and MC simulation. The background shape is parametrized as a linear function. As an example, Fig.~\ref{fig:fit_at_3686} shows the fit results to the selected candidates of the $\psi(3686)$ data sample.

The detection efficiency is determined with the signal MC samples generated based on the PWA results at $\sqrt{s}=3.686$ GeV and 3.773 GeV. The ISR effect is included with {\sc conexc}, which depends on the Born cross sections as input. Therefore, several rounds of iteration are performed. First, the signal MC samples are generated by using the Born cross sections with continuum process only. Then, the detection efficiency $\epsilon$ is obtained and the observed cross section~$\sigma_{\rm obs}$ is calculated with
\begin{equation}
{\sigma_{\rm obs}=\frac{N_{\rm obs}}{\mathcal{L}_{\rm int}\times\epsilon\times \mathcal{B}(\pi^0(\eta)\to \gamma\gamma)}}.
\end{equation}
Next, a fit is performed to the energy-dependent observed cross sections with statistical uncertainty only. The detail of the fit is described in Sec.~\ref{Xobs_fit}. The fitted line shape of the cross sections is used as the input to generate signal MC samples in the next iteration. The process is repeated until the result is stable. Table~\ref{tab:psipscan_Xsection_points} shows the obtained observed cross sections.

\begin{table*}[htbp]
	\tabcolsep=0.3cm
	\caption{The measured observed cross sections of $e^+e^-\to p\bar{p}\pi^0$ and $e^+e^-\to p\bar{p}\eta$ around the $\psi(3686)$ peak. Here, $\mathcal{L}$ is the integrated luminosity; $N_{\rm sig}$ is the number of observed signal event; $\epsilon$ is the detection efficiency; The uncertainties are statistical only.}
	\centering
	\begin{tabular}{cccccccc}
		\hline
		\hline
        $\sqrt{s}$~(MeV) & $\mathcal{L}$~(pb$^{-1}$) & $N^{\pi^0}_{\rm sig}$ & $\epsilon^{\pi^0}$~(\%)  & $\sigma^{\pi^0}_{\rm obs}$~(pb) & $N^{\eta}_{\rm sig}$ & $\epsilon^{\eta}$~(\%)  & $\sigma^{\eta}_{\rm obs}$~(pb)\\
		\hline
		3670.16 & 84.7 & $294\pm18$ & 43.47 &  $8.1\pm0.5$   & $55\pm8$ & 40.96 & $4.0\pm0.6$ \\
		\hline
		3680.14 & 84.8 & $257\pm17$ & 42.56 &  $7.2\pm0.5$   & $41\pm7$ & 40.30 & $3.0\pm0.5$\\
		\hline
		3682.75 & 28.7 & $122\pm12$ & 40.46 &  $10.7\pm1.0$ & $25\pm6$ & 39.63 &$5.6\pm1.2$ \\
		\hline
		3684.22 & 28.7 & $431\pm21$ & 38.42 &  $39.6\pm1.9$ & $96\pm12$ & 39.55 &$21.4\pm2.5$ \\
		\hline
		3685.26 & 26.0 & $984\pm33$ & 38.55 & $99.4\pm3.2$ & $173\pm15$ & 39.65 & $42.6\pm3.7$ \\
		\hline
		3686.10 & 3877.1 & $183914\pm441$ & 38.51 &  $124.7\pm0.3$ & $29920\pm205$ & 39.67 & $49.4\pm0.3$ \\
		\hline
		3686.50 & 25.1 & $1104\pm34$ & 38.51 &  $115.5\pm3.5$ & $201\pm18$ & 39.60 & $51.3\pm4.4$ \\
		\hline
		3691.36 & 69.4 & $504\pm23$ & 40.40 &   $18.2\pm0.8$  & $86\pm11$ & 39.49 &  $8.0\pm1.0$ \\
		\hline
		3709.76 & 70.3 & $294\pm18$ & 38.29 &  $11.1\pm0.6$  & $49\pm8$ & 36.72 & $4.8\pm0.7$ \\
		\hline
		\hline
	\end{tabular}
	\label{tab:psipscan_Xsection_points}
\end{table*}

\section{SYSTEMATIC UNCERTAINTIES}

\subsection{Uncertainties on observed cross sections}~\label{sec:SysU_Xobs}

The systematic uncertainties on the observed cross section measurement include several sources, as summarized in Table~\ref{tab:Sys_tot_Xobs}. They are estimated as described below.

\begin{table*}[htbp]
	\tabcolsep=0.22cm
	\centering
	\caption{Systematic uncertainties on the measurement of observed cross sections of $\psi(3686)\to p\bar{p}\pi^0$ and $\psi(3686)\to p\bar{p}\eta$ in unit of percent.}
	\begin{tabular}{l|ccccccccc}
		\hline
		\hline
		$\psi(3686)\to p\bar{p}\pi^0$  & 3670.16 & 3680.14 & 3682.75 & 3684.22 & 3685.26 & 3686.10 & 3686.50 & 3691.36 & 3709.76\\
		\hline
		Photon detection      & \multicolumn{9}{c}{2.0}\\ \hline
		PID of proton      & \multicolumn{9}{c}{3.0}\\ \hline
		Tracking & \multicolumn{9}{c}{2.0}\\ \hline
		Luminosity & \multicolumn{9}{c}{1.0}\\ \hline
		Helix parameter   & 0.8 & 0.5 &0.7 &0.5 & 0.6 & 0.5 & 0.5 &0.6 & 0.9 \\  \hline
		Peaking background   & \multicolumn{9}{c}{0.3}\\ \hline
		$J/\psi$ veto   & 0.2 &0.4&0.5&0.3&0.1 &$<0.1$&0.1&0.3&0.1 \\  \hline
		$\mathcal{B}$($\pi^0\to\gamma\gamma$)   & \multicolumn{9}{c}{$<0.1$} \\  \hline
		MC modeling   & 0.4 & 0.3 & 0.2 & 0.2 & 0.6 & 0.2 & 0.3 & 0.2 & 0.1 \\  \hline
		Fit to $m_{\gamma\gamma}$ & 1.1 & 0.2 & 0.2 & 0.7 & 0.2 & 0.4 & 0.4 & 0.4 & 0.6 \\  \hline		
		Total                   & 4.5 & 4.3 & 4.3 & 4.4 & 4.3 & 4.3 &4.3 &4.3 &4.4\\
		\hline
		\hline
		$\psi(3686)\to p\bar{p}\eta$  & 3670.16 & 3680.14 & 3682.75 & 3684.22 & 3685.26 & 3686.10 & 3686.50 & 3691.36 & 3709.76\\
		\hline
		Photon detection      & \multicolumn{9}{c}{2.0}\\ \hline
		Tracking & \multicolumn{9}{c}{2.0}\\ \hline
		Luminosity & \multicolumn{9}{c}{1.0}\\ \hline
		Helix parameter   & 0.6 & 0.6 &0.5 &0.5 & 0.5 & 0.6 & 0.4 &0.5 & 0.6 \\  \hline
		Peaking background   & \multicolumn{9}{c}{1.9}\\ \hline
		$J/\psi$ veto   & 0.2&0.2&0.4&0.8 & 0.3 & $<0.1$ &0.2&1.1&0.4 \\  \hline
		$\mathcal{B}$($\eta\to\gamma\gamma$)   & \multicolumn{9}{c}{0.5} \\  \hline
		MC modeling  & 0.1 & 0.8 & 1.1 & 0.4 & 0.1 & 0.3 & 0.1 & 0.2 & $<0.1$ \\  \hline
		Fit to $m_{\gamma\gamma}$ & 4.4 & 2.2 & 4.7 & 1.4 & 2.4 & 0.4 & 2.4 & 6.3 & 17.2 \\  \hline		
		Total                   & 5.7 & 4.3 & 6.0 & 4.0 & 4.4 & 3.7 & 4.3 &7.4&17.6\\
		\hline
		\hline
	\end{tabular}
	\label{tab:Sys_tot_Xobs}
\end{table*}

The integrated luminosity is measured using Bhabha scattering events, with an uncertainty less than 1.0\%~\cite{cited_luminosity}. The uncertainty related to tracking efficiency of proton is estimated to be 1.0\% using a control sample of $J/\psi\to p\pi^-\bar{p}\pi^+$~\cite{proton_sys}. The uncertainty due to the PID efficiency is studied with a control sample of $J/\psi\to pK^-\bar{\Lambda}+c.c.$~\cite{proton_sys}, which is taken to be 1.0\% for proton and 2.0\% for anti-proton, respectively. The uncertainty from photon detection is taken to be 1.0\% per photon~\cite{photon_sys}.

The systematic uncertainty due to the $J/\psi$ veto requirement is estimated by varying the cut range within [0.044, 0.056] GeV with a step size of 0.002 GeV. The difference between the average of re-measured branching fractions with varied cuts and nominal value is taken to be the uncertainty. The systematic uncertainty associated with the kinematic fit is estimated by performing corrections on the track helix parameter in the MC simulation~\cite{helix_sys}. The difference between the detection efficiencies obtained with and without the helix parameter correction is taken as the corresponding systematic uncertainty.

The systematic uncertainty of MC modeling is estimated with a new signal MC sample, in which all fitted complex coupling constants, quoted resonance parameters are smeared with their uncertainties. The difference of the detection efficiencies obtained with the new and nominal MC samples is taken as the uncertainty. The systematic uncertainty from the fit to the $m_{\gamma\gamma}$ spectrum is taken into account in two aspects. The uncertainty associated with the fit range is estimated by varying the fit range by 5 MeV. The uncertainty from the background shape is estimated by using a second-order polynomial in the fit. For each of these two effects, the difference between the signal yields obtained with nominal and alternative fit is taken as the corresponding systematic uncertainty. Adding these two items in quadrature, we obtain the combined systematic uncertainty from these two effects. The systematic uncertainty due to peaking background is estimated by studying inclusive MC sample. The systematic uncertainty due to the quoted branching fractions of $\mathcal{B}(\pi^0\to\gamma\gamma)$ and $\mathcal{B}(\eta\to\gamma\gamma)$ are $<0.1$\% and $0.5$\%~\cite{PDG}, respectively.

\subsection{Uncertainties on intermediate resonance measurements}

The systematic uncertainties on the intermediate resonance measurements are divided into two categories. The first one is non-PWA related, including photon detection, tracking efficiency of proton, helix parameter, normalization factor $f_c$, interference with continuum process, $\mathcal{B}$($\pi^0\to\gamma\gamma$) or $\mathcal{B}$($\eta\to\gamma\gamma$), and the number of $\psi(3686)$. The overlapping uncertainties are estimated in the same way as in Sec.~\ref{sec:SysU_Xobs}.

The systematic uncertainty due to the normalization factor $f_c$ is estimated by varying $f_c$ within its uncertainty. The difference between signal yields obtained with nominal and alternative $f_c$ is taken to be the uncertainty. The systematic uncertainty caused by an unknown phase angle between the $\psi(3686)$ resonance and continuum process is estimated by following Ref.~\cite{interference_sys}. The ratio of the maximum impact from the interference term with respect to the resonance term is defined as
\begin{equation}
r^{\rm max}_{R}=\frac{2}{\hbar c}AB\sin(\pi/2), A=\sqrt{\frac{\sigma_c}{\mathcal{B}}}.
\end{equation}
Here, $\sigma_c$ is the cross section of the continuum process $e^+e^-\to N^* \bar{p}+c.c.\to p\bar{p}\pi^0(\eta)$ and $\mathcal{B}$ is the branching fraction of $\psi(3686)\to N^*\bar{p}+c.c.\to p\bar{p}\pi^0(\eta)$. The factor $B=M_{\psi(3686)}/\sqrt{12\pi\mathcal{B}(\psi(3686)\to e^+e^-)}=6.74$~GeV is a constant depending on the $\psi(3686)$ resonance parameters. To be conservative, $r^{\rm max}_{R}$ is taken to be the uncertainty. The systematic uncertainty on the total number of $\psi(3686)$ events is 0.5\%.

The PWA related systematic uncertainties include the used mass and width, energy-dependent width parameters, the $J/\psi$ veto requirement of $|M_{p\bar{p}}-3.097|<0.05$ GeV, hadronic background, and fit bias. The systematic uncertainty due to hadronic background is estimated by rescaling the background level according to our study on the data sample. The systematic uncertainty related to extra resonance states is studied by adding $N^*$ states with less experimental confirmation or higher spin, {\it i.e.} $N(1675)$, $N(1680)$, and $N(1700)$ for $\psi(3686)\to p\bar{p}\pi^0$ but none for $\psi(3686)\to p\bar{p}\eta$, into our solution. Although the statistical significance for each additional resonance is greater than 5$\sigma$, the corresponding fit fractions are all less than 3\%. Therefore no additional resonances are included in the fit, but are accounted as sources of systematic uncertainty. The states with higher spin or weaker experimental confirmation are not considered in this work. The systematic uncertainty due to the energy-dependent width parameters is estimated by varying the branching fractions $r_i$ in Table~\ref{tab:Rparas} within their individual uncertainties. For the above three sources, the maximum difference of signal yield obtained with the nominal and alternative fit is taken to be their uncertainty. The systematic uncertainty caused by the requirement of $|M_{p\bar{p}}-3.097|<0.05$ GeV is studied in the same way as in Sec.~\ref{sec:SysU_Xobs}. The systematic uncertainty due to the fixed masses and widths in the nominal fit, is estimated by varying the fixed values according to their uncertainties, shown in Table~\ref{tab:baseline}, within 3$\sigma$ region, and repeating the fit. In total, 40 fits are performed with mass and width variations and the standard deviations of the measured branching fractions are taken as the associated systematic uncertainty. The systematic uncertainty due to fit bias is studied by performing an input/output check. Toy MC samples are generated according to the PWA fit result with simulated background events added. The same PWA fit is performed to each toy MC sample and the pull distributions of fitted signal yields are obtained. Any non-zero mean value of a pull distribution, indicating a possible fit bias, is taken as its uncertainty.

The systematic uncertainties for $\psi(3686)\to p\bar{p}\pi^0$ and $\psi(3686)\to p\bar{p}\eta$ are summarized in Tables~\ref{tab:Sys_pi0_all} and~\ref{tab:Sys_eta}, respectively. 

  \begin{table*}[htbp]
	\caption{Relative systematic uncertainties of the branching fractions for intermediate resonance in $\psi(3686)\to p\bar{p}\pi^0$ (in percent).}
	\centering
	\tabcolsep=0.18cm
	\scriptsize
	\begin{tabular}{l|c|c|c|c|c|c|c|c|c|c}
		\hline
		\hline
		Source  & $N(940)$ & $N(1440)$ & $N(1520)$ & $N(1535)$ & $N(1650)$ & $N(1710)$ & $N(1720)$ & $N(2100)$ & $N(2300)$ & $N(2570)$\\
		\hline
		Photon detection      & \multicolumn{10}{c}{2.0} \\ \hline
		Tracking  & \multicolumn{10}{c}{2.0} \\  \hline
		Helix parameter   & 0.6 & 0.7 & 0.6 & 0.5 & 0.5 & 0.6 & 0.7 & 0.5 & 0.6 & 0.5 \\  \hline
		Normalization factor $f_c$& $<0.1$&$<0.1$&0.1&$<0.1$&$<0.1$&0.1&$<0.1$&$<0.1$&0.2&0.1\\  \hline
		Interference with continuum& 5.2&9.1&24.9&8.0&11.9&20.3&12.8&17.4&34.0&23.7\\  \hline
		$\mathcal{B}$($\pi^0\to\gamma\gamma$)   & \multicolumn{10}{c}{$<0.1$} \\  \hline
		Number of $\psi(3686)$& \multicolumn{10}{c}{0.5} \\  \hline
		\hline
		Quoted mass and width  & 6.6&11.4&16.9&13.3&24.7&47.7&36.1&30.8&66.5&29.7 \\  \hline
		Extra resonance states  & 2.4&20.5&4.4&12.6&20.6&77.4&7.2&26.3&30.2&63.9 \\  \hline
		Energy-dependent width&
		1.3 & 5.5 & 0.7 & 3.9 & 1.8 & 10.2 & 1.3 & 1.1 & 6.2 & 1.4\\  \hline
		Hadronic background & 3.8&1.9&0.1&0.9&0.4&3.4&0.5&0.7&7.7&0.6 \\  \hline
		$J/\psi$ veto  &2.9 & 0.1&0.8&1.4&1.3&3.2&2.5&1.9&7.7 &2.5 \\  \hline
		Fit bias &  0.7&1.4&1.2&0.3&2.0&7.6&1.8&8.9&6.2&2.3\\  \hline		
		Total                   & 10.5&26.0&30.6&20.6&34.6&94.2&39.2&45.1&81.8&74.5 \\
		\hline
		\hline
	\end{tabular}
	\label{tab:Sys_pi0_all}
\end{table*}

\begin{table*}[htbp]
	\caption{Relative systematic uncertainties of the branching fractions for $N(1535)$ and $N(1710)$ in $\psi(3686)\to p\bar{p}\eta$ (in percent).}
	\centering
	\begin{tabular}{l|cc}
		\hline
		\hline
		Source  & $N(1535)$ & $N(1710)$\\
		\hline
		Photon detection      & \multicolumn{2}{c}{2.0}   \\ \hline
		Tracking & \multicolumn{2}{c}{2.0}\\  \hline
		Helix parameter   & 0.7 & 0.5  \\  \hline
		Normalization factor $f_c$& $<0.1$ &  $<0.1$ \\ \hline
		Interference with continuum& 13.3 & 6.9 \\  \hline
		$\mathcal{B}$($\eta\to\gamma\gamma$)   & \multicolumn{2}{c}{$0.5$}  \\  \hline
		Number of $\psi(3686)$& \multicolumn{2}{c}{0.5}  \\  \hline
		\hline
		Quoted mass and width  & 2.9 & 39.0  \\  \hline
		Extra resonance states& \multicolumn{2}{c}{...}  \\  \hline
		Running width parameters& 0.3 & $<0.1$ \\  \hline
		Hadronic background & 0.5 & 1.2 \\  \hline
		$J/\psi$ veto   & 0.7 & 0.4 \\  \hline
		Fit bias & 0.7 & 4.8 \\ \hline		
		Total                   & 14.0 & 40.0 \\
		\hline
		\hline
	\end{tabular}
	\label{tab:Sys_eta}
\end{table*}

\section{FIT TO OBSERVED CROSS SECTIONS}~\label{Xobs_fit}
Least square fits are performed to the observed cross sections of $e^+e^-\to p\bar{p}\pi^0$ and $e^+e^-\to p\bar{p}\eta$ to obtain the branching fractions of $\psi(3686)\to p\bar{p}\pi^0$ and $\psi(3686)\to p\bar{p}\eta$, respectively.

The $\chi^2$ is constructed as
\begin{equation}
\chi^2= (\Delta\vec{\sigma})^T V^{-1} \Delta\vec{\sigma},
\end{equation}
where the difference in cross section $\Delta\vec{\sigma}_{i}=\sigma_{i}-\sigma^{\rm fit}_{i}(\vec{\theta})$ and $V$ is the covariance matrix. The $\sigma_{i}$ and $\sigma^{\rm fit}_{i}$ are the measured and fitted values for cross section at the $i$-th energy point, respectively. The covariance matrix is constructed as $V_{ii}=\sigma^2_{{\rm sta},i}+\sigma^2_{{\rm sys-corr},i}+\sigma^2_{{\rm sys-uncorr},i}+(\sigma_{{\rm beam},i}\times \frac{d\sigma_{\rm obs}}{d\sqrt{s}})^2$ for diagonal elements and $V_{ij}=\sigma_{ {\rm sys-corr},i}\times\sigma_{{\rm sys-corr},j}$ for off-diagonal elements~($i\neq j$). The $\sigma_{\rm sys-corr}$ includes the systematic uncertainties of tracking, PID, and $\mathcal{B}(\pi^0\to\gamma\gamma)$ or $\mathcal{B}(\eta\to\gamma\gamma)$. The $\sigma_{{\rm beam},i}=0.3$ MeV is the uncertainty on the beam energy.

The line shape of observed cross sections $\sigma_{\rm obs}(s)$ is constructed as follows. The Born cross section is modeled with
\begin{equation}
\sigma_{\rm Born}=|A_{\rm con}+A_{\rm res}\times e^{i\phi}|^2.
\end{equation}
The continuum contributions $A_{\rm con}(s)=a/s^n$ are fixed to those in Refs.~\cite{exp_ee2pppi0} and~\cite{exp_ee2ppeta} for $\pi^0$ and $\eta$ channels, respectively. The $A_{\rm res}$ is defined as
\begin{equation}
A_{\rm res}(s) = \frac{\sqrt{12\pi\Gamma_{ee}\Gamma_{\rm tot}\mathcal{B}_{f}}}{s-M^2+iM\Gamma_{\rm tot}}.
\end{equation}
Here, $\Gamma_{ee}$ is the electronic partial width; $\Gamma_{\rm tot}$ is the total width; $\mathcal{B}_{f}$ is the branching fraction of $\psi(3686)\to p\bar{p}\pi^0$ or $\psi(3686)\to p\bar{p}\eta$; $M$ is the mass of $\psi(3686)$. In the fit, $\Gamma_{\rm tot}$, and $M$ are fixed to the PDG values~\cite{PDG}. The impact from ISR effect on the cross section is included by 
\begin{equation}
\sigma^{\prime}(s)=\int^{\sqrt{s}}_{M_{\rm cut}}dm\frac{2m}{s}W(s,x)\sigma_{\rm Born}(m),
\end{equation}
where $W(s,x)$ is the radiative function up to $\mathcal{O}(\alpha^2)$~\cite{conexc} and $M_{\rm cut}$ is set to be 3.65 GeV. This integration is performed in the same way as in Ref.~\cite{conexc}. The beam spread effect is considered by performing a Gaussian convolution to the cross section with
\begin{equation}
\sigma_{\rm obs}(s)=\int ds^\prime\sigma^{\prime}(s)\frac{1}{\sqrt{2\pi}S_{E}}\exp\left(\frac{-(s-s^\prime)^2}{2S^2_E}\right),
\end{equation}
where the $S_{E}$ is the beam energy spread.

\begin{figure*}[htbp]
	\centering
	\begin{minipage}[t]{0.45\textwidth}
		\includegraphics[width=8.5cm]{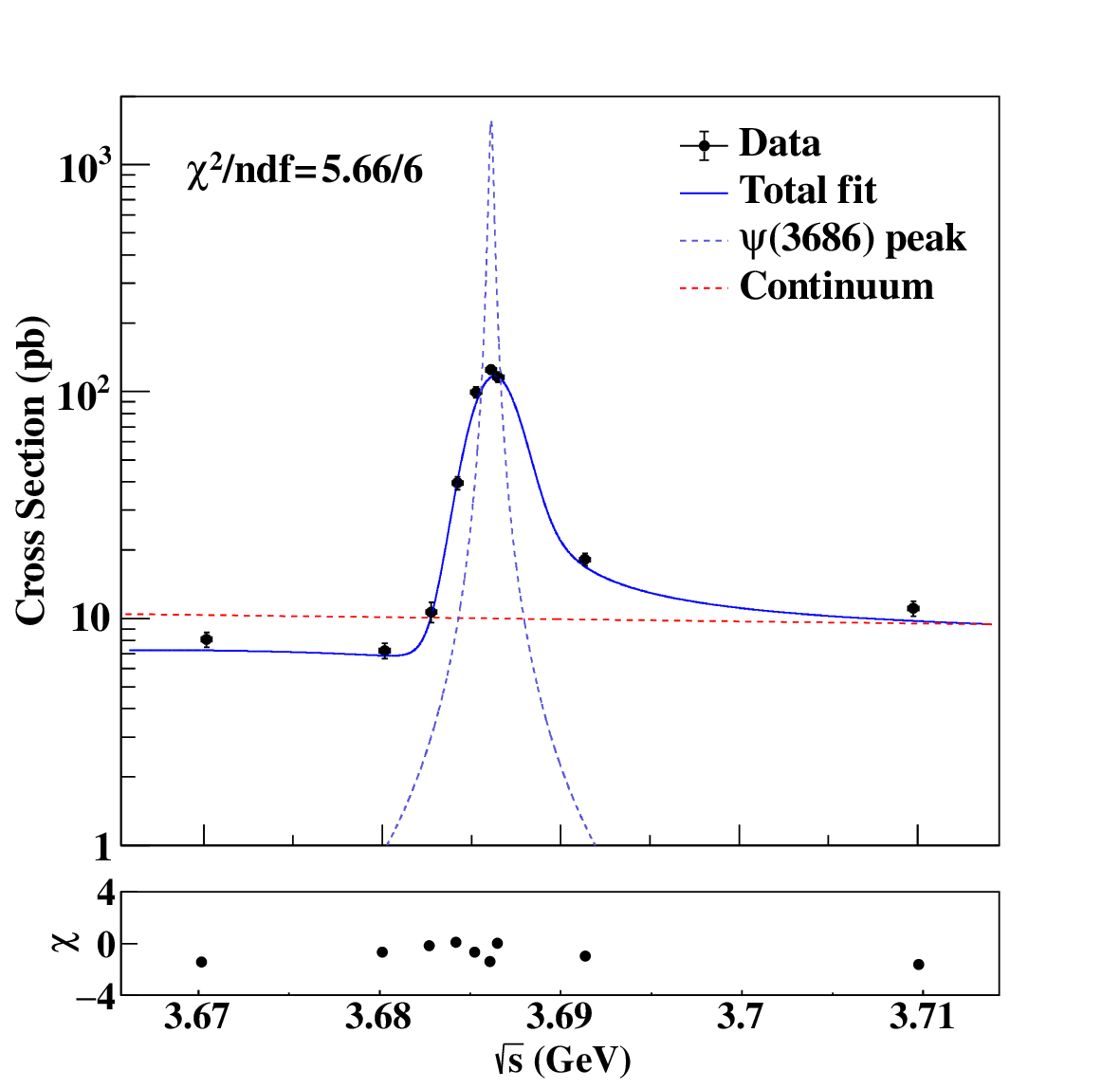}
	\end{minipage}
	\begin{minipage}[t]{0.45\textwidth}
		\includegraphics[width=8.5cm]{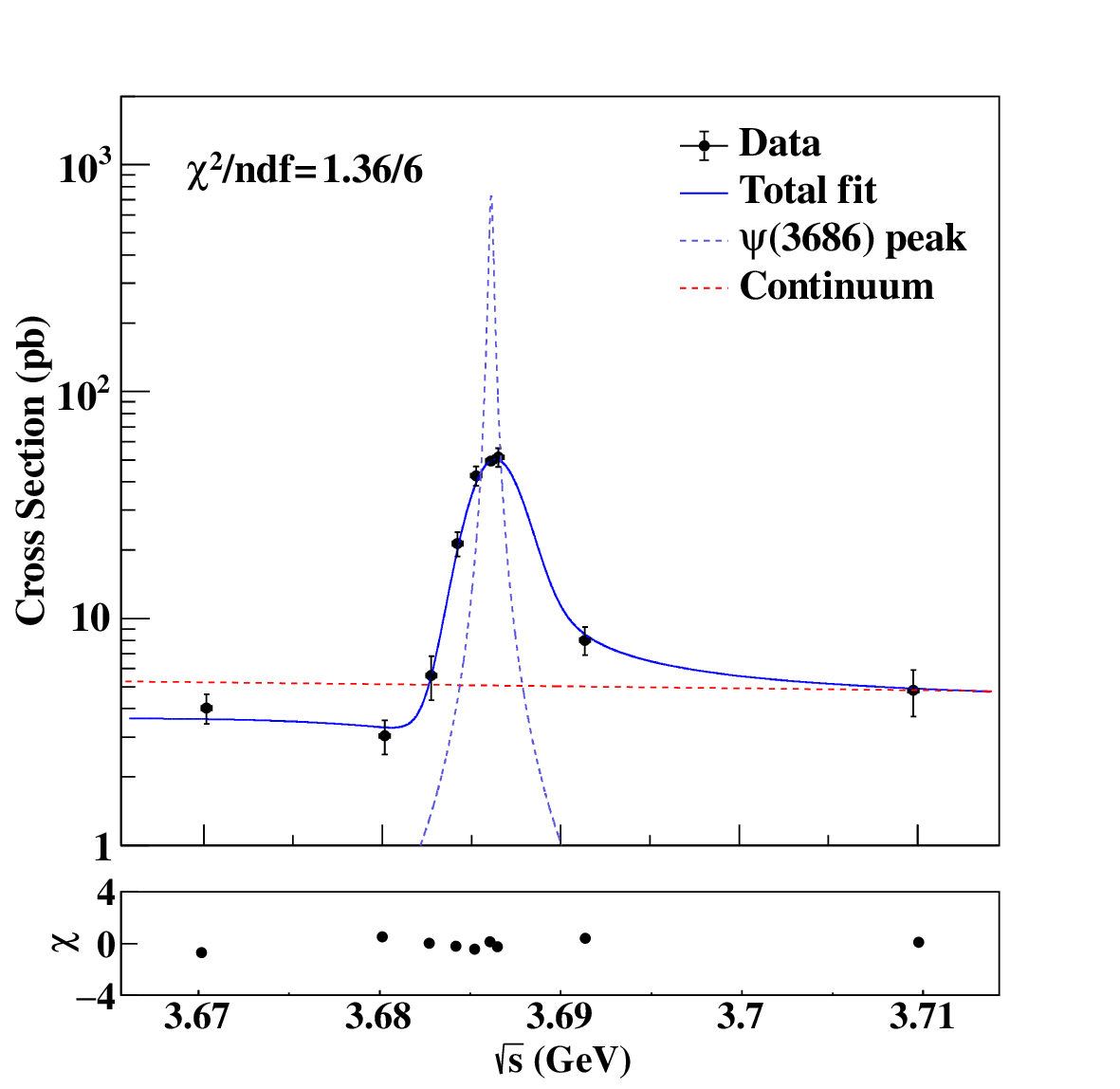}
	\end{minipage}
	\caption{Fit to the observed cross sections of (a) $\psi(3686)\to p\bar{p}\pi^0$ and (b) $\psi(3686)\to p\bar{p}\eta$ around the $\psi(3686)$ peak. Destructive solutions are shown as examples. The dots with error bars are data, the blue solid lines are the fit results, the cyan dashed lines are the $\psi(3686)$ peaks without beam spread effect, and the red dashed lines are the continuum processes.}
	\label{fig:fit_Xobs}
\end{figure*}

Figure~\ref{fig:fit_Xobs} shows the fit results to $\psi(3686)\to p\bar{p}\pi^0$ and $\psi(3686)\to p\bar{p}\eta$. The fitted parameters and calculated branching fractions are summarized in Table~\ref{tab:psipscan_fit_par}.

\begin{table*}[htbp]
	\tabcolsep=0.25cm
	\caption{Fitted parameters of $\psi(3686)\to p\bar{p}\pi^0$ and $\psi(3686)\to p\bar{p}\eta$, and corresponding branching fractions $\mathcal{B}_{f}$. The first uncertainties include both statistical and systematic. The second uncertainties of $\mathcal{B}_{f}$ are from $\Gamma_{ee}=(2.33\pm0.04)$ KeV~\cite{PDG}.}
	\centering
	\begin{tabular}{ccccc}
		\hline
		\hline
		$\psi(3686)\to p\bar{p}\pi^0$ & $\mathcal{B}_{f}\Gamma_{ee}$ (0.1~eV) & $\phi$ $(^\circ)$ & $\Delta E$ (MeV)  & $\mathcal{B}_{f}$ $(\times10^{-6})$ \\
		\hline
		Constructive solution & $3.12\pm0.26$ & $65.0\pm6.7$ & $1.27\pm0.09$ & $133.9\pm11.2\pm2.3$ \\
		\hline
		Destructive solution & $4.28\pm0.32$ & $-68.9\pm5.7$ & $1.27\pm0.09$ & $183.7\pm13.7\pm3.2$ \\
		\hline
		\hline
		$\psi(3686)\to p\bar{p}\eta$ & $\mathcal{B}_{f}\Gamma_{ee}$ (0.1~eV) & $\phi$ $(^\circ)$ & $\Delta E$ (MeV)  & $\mathcal{B}_{f}$ $(\times10^{-6})$ \\
		\hline
		Constructive solution & $1.44\pm0.15$ & $58.9\pm14.1$ & $1.39\pm0.14$ & $61.5\pm6.5\pm1.1$ \\
		\hline
		Destructive solution & $1.98\pm0.16$ & $-63.8\pm12.1$ & $1.39\pm0.14$ & $84.4\pm6.9\pm1.4$ \\
		\hline
		\hline
	\end{tabular}
	\label{tab:psipscan_fit_par}
\end{table*}

\section{RATIO OF $\Gamma_{N(1535)\to N\eta}/\Gamma_{N(1535)\to N\pi}$}

The ratio of decay widths $\Gamma_{N(1535)\to N\eta}/\Gamma_{N(1535)\to N\pi}$ does not depend on whether $N(1535)\bar{p}+c.c.$ is from $\gamma^*$ or $\psi(3686)$. Therefore, the systematic uncertainties from the interference between $\psi(3686)$ resonance and continuum process and normalization factor $f_c$, and some common terms, including photon detection, tracking, total number of $\psi(3686)$ events, and helix parameter, are expected to cancel in the ratio and are therefore excluded.

The ratio of the partial decay widths between $N(1535)\to N\eta$ and $N(1535)\to N\pi$ is calculated by
\begin{equation}
\frac{\Gamma_{N(1535)\to N\eta}}{\Gamma_{N(1535)\to N\pi}}=\frac{N^{\eta}_{\rm sig}/(\epsilon^{\eta}\times\mathcal{B}(\eta\to\gamma\gamma))}{3\times N^{\pi^0}_{\rm sig}/(\epsilon^{\pi^0}\times\mathcal{B}(\pi^0\to\gamma\gamma))},
\end{equation}
where the factor 3 includes both $p\pi^0$ and $n\pi^{+}+c.c.$ contributions by assuming isospin conservation. Finally, $\Gamma_{N(1535)\to N\eta}/\Gamma_{N(1535)\to N\pi}$ is determined to be $0.99\pm 0.05\pm0.19$, where the first uncertainty is statistical and the second one is systematic.

\section{SUMMARY}

In summary, using a sample of $(2712\pm14)\times10^6$ $\psi(3686)$ events collected with the BESIII detector, partial wave analyses of the decays $\psi(3686)\to p\bar{p}\pi^0$ and $\psi(3686)\to p\bar{p}\eta$ are performed.  The data can be well described by considering several well-established $N^*$ states.

Taking into account the interference effects in this analysis, the improved measurement on branching fractions of $\psi(3686)\to p\bar{p}\pi^0$ and $\psi(3686)\to p\bar{p}\eta$ are presented. The deviations from previous BESIII results~\cite{exp_2,exp_3} are attributed mainly to the absence of consideration of interference effects in the previous measurements. Combining $\mathcal{B}(J/\psi\to p\bar{p}\pi^0)$ from the PDG~\cite{PDG}, the ratio of $\mathcal{B}(\psi(3686)\to p\bar{p}\pi^0)/\mathcal{B}(J/\psi\to p\bar{p}\pi^0)$ is determined to be $(11.3\pm1.2)\%$ and $(15.4\pm1.6)\%$, for constructive and destructive interference, respectively. Combining $\mathcal{B}(J/\psi\to p\bar{p}\eta)$ from the PDG~\cite{PDG}, the ratio of $\mathcal{B}(\psi(3686)\to p\bar{p}\eta)/\mathcal{B}(J/\psi\to p\bar{p}\eta)$ is determined to be $(3.1\pm0.4)\%$ and $(4.2\pm0.4)\%$, for constructive and destructive interferences, respectively. The ``$12\%$ rule"~\cite{theory_12p_1,theory_12p_2,theory_12p_3} is significantly violated in the $p\bar{p}\eta$ final state but is confirmed in the $p\bar{p}\pi^0$ final state.  The branching fractions of several $N^*$ intermediate states are determined as well in this analysis.

The ratio of $\Gamma_{N(1535)\to N\eta}/\Gamma_{N(1535)\to N\pi}$ is determined to be $0.99\pm 0.05\pm0.19$, which is consistent with $\Gamma_{N(1535)\to N\eta}/\Gamma_{N(1535)\to N\pi}=1.00\pm 0.40$ averaged by the PDG~\cite{PDG} based on fixed-target experiment results and previous BESIII publications~\cite{exp_2,exp_3}, but with much improved precision. This result confirms the exotic property of $N(1535)$ in an $e^+e^-$ collider experiment for the first time and suggests a strong $s\bar{s}$ component in $N(1535)$~\cite{flavor_sym}. 

\section{ACKNOWLEDGMENTS}

The BESIII collaboration thanks the staff of BEPCII and the IHEP computing center for their strong support. This work is supported in part by National Key R\&D Program of China under Contracts Nos. 2020YFA0406300, 2020YFA0406400; National Natural Science Foundation of China (NSFC) under Contracts Nos. 12075252, 12175244, 11875262, 11635010, 11735014, 11835012, 11935015, 11935016, 11935018, 11961141012, 12022510, 12025502, 12035009, 12035013, 12192260, 12192261, 12192262, 12192263, 12192264, 12192265; the Chinese Academy of Sciences (CAS) Large-Scale Scientific Facility Program; Joint Large-Scale Scientific Facility Funds of the NSFC and CAS under Contract No. U1832207; CAS Key Research Program of Frontier Sciences under Contract No. QYZDJ-SSW-SLH040; 100 Talents Program of CAS; INPAC and Shanghai Key Laboratory for Particle Physics and Cosmology; ERC under Contract No. 758462; European Union's Horizon 2020 research and innovation programme under Marie Sklodowska-Curie grant agreement under Contract No. 894790; German Research Foundation DFG under Contracts Nos. 443159800, Collaborative Research Center CRC 1044, GRK 2149; Istituto Nazionale di Fisica Nucleare, Italy; Ministry of Development of Turkey under Contract No. DPT2006K-120470; National Science and Technology fund; National Science Research and Innovation Fund (NSRF) via the Program Management Unit for Human Resources \& Institutional Development, Research and Innovation under Contract No. B16F640076; STFC (United Kingdom); Suranaree University of Technology (SUT), Thailand Science Research and Innovation (TSRI), and National Science Research and Innovation Fund (NSRF) under Contract No. 160355; The Royal Society, UK under Contracts Nos. DH140054, DH160214; The Knut and Alice Wallenberg Foundation, Sweden, the Swedish Research Council, the Swedish Foundation for International Cooperation in Research and Higher Education (STINT); U. S. Department of Energy under Contract No. DE-FG02-05ER41374.

\end{document}